\documentclass[showpacs,aps,amsmath,amssymb,pre,preprint,preprintnumbers]{revtex4}
\newcommand{\BE}{\begin{equation}}
\newcommand{\EE}{\end{equation}}
\newcommand{\BA}{\begin{eqnarray}}
\newcommand{\EA}{\end{eqnarray}}
\usepackage{graphicx}% Include figure files
\usepackage{dcolumn}% Align table columns on decimal point
\usepackage{bm}% bold math
\input epsf.tex
\begin{document}

\title{Electromagnetic energy and energy flows in photonic crystals made of arrays of parallel dielectric cylinders}

\author{Chao-Hsien Kuo}\author{Zhen
Ye}\email{zhen@phy.ncu.edu.tw} \affiliation{Wave Phenomena
Laboratory, Department of Physics, National Central University,
Chungli, Taiwan 32054, Republic of China}

\date{\today}

\begin{abstract}

We consider the electromagnetic propagation in two-dimensional
photonic crystals, formed by parallel dielectric cylinders
embedded a uniform medium. The frequency band structure is
computed using the standard plane-wave expansion method, and the
corresponding eigne-modes are obtained subsequently. The optical
flows of the eigen-modes are calculated by a direct computation
approach, and several averaging schemes of the energy current are
discussed. The results are compared to those obtained by the usual
approach that employs the group velocity calculation. We consider
both the case in which the frequency lies within passing band and
the situation in which the frequency is in the range of a partial
bandgap. The agreements and discrepancies between various
averaging schemes and the group velocity approach are discussed in
detail. The results indicate the group velocity can be obtained by
appropriate averaging method.

\end{abstract}

 \maketitle

\section{Introduction}

When propagating through periodically structured media, i.~e.
photonic crystals (PCs), optical waves will be modulated by the
periodicity. As a result, the dispersion of waves will no longer
behave as that in a free space, and frequency band structures will
appear. Under certain conditions, waves may be prohibited from
propagation in certain or all directions, corresponding to partial
and complete bandgaps respectively. The photonic crystals which
could reveal bandgaps are called bandgap materials.

Photonic crystals (PCs) are usually made of periodically
structured materials which are sensitive to electromagnetic waves,
and have been studied both intensively and
extensively\cite{Yariv,Book}. The research of waves in periodic
media has been first put on a theorematic basis stemmed from the
efforts of Bloch\cite{Bloch} and Brillouin\cite{Bri} towards
electronic systems. An early survey was given in the excellent
textbook\cite{Yariv,Bri2}. The general aspects of electromagnetic
waves in photonic crystals are furthered reviewed in
Refs.~\cite{Yariv,Book}. A comprehensive survey of the literature
can be referred to Ref.~\cite{web}.

The propagation of electromagnetic waves in crystal structures is
one of the top issues in the research of photonic crystals. The
common theoretical approach to electromagnetic propagation in
periodic media has been given in Ref.~\cite{Yariv}, and may be
summarized as follows. The Maxwell equations are first derived for
waves in periodic media. By Bloch theorem, the solution can be
expanded in terms of plane waves. The solution is then substituted
into the governing equations to obtain an eigen-equation that
determines the dispersion relations between the frequency and the
wave vector that lies within the first Brillouin zone. These
relations are termed as frequency band structures. Since it has
been proved\cite{Yariv,Yeh} that the averaged energy velocity
equals the group velocity which can be obtained as the gradient of
the dispersion relations with the respect to the space of wave
vectors, the investigation of electromagnetic propagation in
periodic structures is thus reduced to the calculation of the
group velocity from the band structures, thereby making a
significant simplification of the problem. This approach may be
termed as the group velocity approach and will be denoted as GVA
hereafter.

Since the average of the energy velocity, represented by the
Poynting vector $\langle\vec{S}\rangle \sim
\langle\vec{E}\times\vec{H}\rangle$, is performed over the whole
unit cell of periodic media, a few questions may be asked about
this group velocity approach. The first question is whether such
averaged energy flow can fully depict the actual electromagnetic
energy flow in periodic media. Second, since in actual
measurements it is often hard to detect the physical quantities
within the periodic media, how to obtain the group velocity
without having to put a detector into the media needs also to be
considered. Deducing the group velocity from practical
measurements is in fact an important task in the photonic crystal
research. More explicitly, the question is how to relate practical
measurements with the group velocity which can be derived
theoretically from the band structure calculation.

The purpose of the present paper is to examine the question of how
well GVA can describe the actual electromagnetic (EM) energy flow
in periodic structures, and discuss the issue how to obtain the
group velocity from practical perspectives. We will compare the
averaged energy flow obtained from GVA with the results from the
direct computation of the energy current. Various schemes are
proposed to obtain the averaged current and compared, so to find
the most appropriate schemes in deducing the group velocity in
practical measurements.

To simplify our discussion yet without losing generality, we will
only consider the propagation of electromagnetic waves in two
dimensional periodic media, i.~e. two dimensional photonic
crystals, which are made of arrays of dielectric cylinders. A
reason of using this type of systems is that the systems are
experimentally ready, and therefore the conclusion derived from
the present paper can be verified. The significance of the present
research is two folds. First, it provides useful information about
how to obtain the group velocity. Second, it shows under what
conditions the group velocity approach is valid.

The paper is structured as follows. In the next section, we will
present the usual formulation of Maxwell's equations for EM waves
in periodic structures. The energy flow will be formulated, and
the GVA will be outlined in general. In section III, the
particular system will be discussed, and a few averaging schemes
of the energy current will be proposed. The numerical results for
a number of situations will be presented in Section IV. The paper
will be concluded by a summary in section V. The proof of the
equivalence between the averaged energy velocity and the group
velocity will be briefly repeated in the Appendix, for the
readers' convenience.

\section{The general formulation}

\subsection{Bloch wave solution, energy current, and energy density}

The EM waves in two dimensional media can be separated into two
cases of polarization: (1) the s polarization or the
E-polarization, that is, the electric field is along the $z$ axis
perpendicular to the plane of propagation, which is the plane of
the periodicity; and (2) the p polarization or the H polarization,
with the magnetic field being perpendicular to the plane of
propagation. Here we outline the determining equations for EM
propagation in two dimensional periodic media. The details can be
referred to in Refs.~\cite{Book,Yariv}

For both polarizations, the governing equation can be unified as
\BE \nabla\left(\frac{1}{\rho}\nabla p(\vec{r})\right) +
\frac{\omega^2}{c^2(\vec{r})q}p(\vec{r}) = 0,\label{eq:wave}\EE
where $c^2$, $\rho$ and $q$ are two dimensional periodic
functions, depending on the properties of the medium. For the s
polarization, $p$ stands for $E_z$, with $\rho(\vec{r})
=\mu(\vec{r})$ and $q = \mu(\vec{r})$. For the p polarization, $p$
denotes the magnetic field $H_z$ perpendicular to the wave
propagation. In this case, $\rho(\vec{r}) = \epsilon(\vec{r})$ and
$q =\epsilon(\vec{r})$.

Due to the periodicity, we can make the following expansions \BE
\frac{1}{\rho} = \sum_{\vec{G}}\sigma(\vec{G})
e^{i\vec{G}\cdot\vec{r}}, \ \mbox{and}\ \frac{1}{c^2q} =
\sum_{\vec{G}}\chi(\vec{G}) e^{i\vec{G}\cdot\vec{r}},
\label{eq:exp1}\EE where $\vec{G}$ are the reciprocal vectors.

By Bloch's theorem, the solution to Eq.~(\ref{eq:wave}) can be
expressed as \BE p_{\vec{K}}(\vec{r}) = e^{i\vec{K}\cdot\vec{r}}
u_{\vec{K}}(\vec{r}),\label{eq:exp2}\EE where $\vec{K}$ is the
Bloch vector that lies within the first Brillouin zone, and
$u_{\vec{K}}$ is a periodic function with the periodicity of the
medium; therefore $p_{\vec{K}}$ is the eigen field corresponding
to the Block vector $\vec{K}$. The function $u_{\vec{K}}$ can be
expanded as \BE u_{\vec{K}}(\vec{r}) = \sum_{\vec{G}}
A_{\vec{K}}(\vec{G})e^{i\vec{G}\cdot\vec{r}}. \label{eq:u}\EE

Substituting Eqs.~(\ref{eq:exp1}) and (\ref{eq:exp2}) into
Eq.~(\ref{eq:wave}), we obtain \begin{widetext}\BE
\sum_{\vec{G}_{1}}\left[ \sigma (\vec{G}_{1})\left( (
\vec{K}+\vec{G}_{2})\cdot (\vec{K}+
\vec{G}_{2}+\vec{G}_{1})\right) -\omega ^{2}\chi (
\vec{G}_{1})\right] A_{\vec{K}}(\vec{G}_{2})=0 \label{unsym}
\EE\end{widetext} From this equation, we can find a secular
equation that determines the dispersion relation between $\omega$
and $\vec{K}$, \begin{widetext}\BE \mbox{det}\left[ \sigma
(\vec{G}_{1})\left( ( \vec{K}+\vec{G}_{2})\cdot (\vec{K}+
\vec{G}_{2}+\vec{G}_{1})\right) -\omega ^{2}\chi (
\vec{G}_{1})\right]_{\vec{G}_1,\vec{G}_2} = 0. \EE
\end{widetext}
Once the dispersion relation is determined, and the
coefficients $A_{\vec{K}}(\vec{G})$ can be obtained from
Eq.~(\ref{unsym}). The EM waves can be subsequently obtained from
Eqs.~(\ref{eq:exp2}) and (\ref{eq:u}).

When either the electrical $E$ or magnetic field $H$ is
determined, corresponding to the s or p polarization respectively,
the magnetic or electric field for the Bloch vector $\vec{K}$ can
be determined from \BE \nabla\times\vec{H}_{\vec{K}} = -i\omega
\epsilon\vec{E}_{\vec{K}}, \ \mbox{or}\
\nabla\times\vec{E}_{\vec{K}} = i\omega \mu\vec{H}_{\vec{K}},
\label{eq:EM}\EE where we have assume $e^{-i\omega t}$ time
dependence.

By Eq.~(\ref{eq:EM}), the time averaged {\it flux} of energy at
any spatial point is subsequently obtained from \BE
\vec{J}_{\vec{K}} = \frac{1}{2}
\mbox{Re}[\vec{E}_{\vec{K}}\times\vec{H}_{\vec{K}}^\star],
\label{eq:J}\EE where `$\star$' refers to the complex conjugate
operation. And the time averaged energy {\it density} is \BE
U_{\vec{K}} = \frac{1}{4}[\epsilon |E_{\vec{K}}|^2 + \mu
|H_{\vec{K}}|^2].\label{eq:U}\EE From these two equations, we can
directly calculate the EM energy flow and density.

\subsection{Group velocity approach (GVA)}\label{GVA}

In principle, the EM propagation in periodic media can be inferred
from a direct calculation using Eqs.~(\ref{eq:J}) and
(\ref{eq:U}). However, it is common to use the group velocity
method to discern the EM energy flows in periodic media. This
approach is based on the following theorem\cite{Yariv}. The group
velocity in periodic media is defined as \BE \vec{v}_g \equiv
\nabla_{\vec{K}}\omega.\EE Or in  another form, \BE \delta \omega
= \vec{v}_g\cdot\delta \vec{K}.\label{eq:vg}\EE The energy
velocity in the periodic media is defined as \BE
\langle\vec{v}_e\rangle = \frac{\frac{1}{V}\int \vec{J}_{\vec{K}}
d^3{r}}{\frac{1}{V}\int U d^3{r}} \equiv \frac{\langle
\vec{J}_{\vec{K}}\rangle}{\langle U_{\vec{K}}\rangle},\EE where
$V$ is the volume of a unit cell, and the integration is performed
in the unit cell. In two dimensions, this equation is reduced to
\BE \langle\vec{v}_e\rangle = \frac{\frac{1}{S}\int
\vec{J}_{\vec{K}} d^2{r}}{\frac{1}{S}\int U_{\vec{K}} d^2{r}}
\equiv \frac{\langle \vec{J}_{\vec{K}}\rangle}{\langle
U_{\vec{K}}\rangle},\label{eq:ve1}\EE with $S$ being the area of a
unit cell and the integration being restricted to the unit cell.
It can be proved\cite{Yariv} that \BE \langle\vec{v}_e\rangle =
\vec{v}_g = \nabla_{\vec{K}}\omega.\label{eq:vgve}\EE This
equation will be referred to as the equivalency theorem.

By Eq.~(\ref{eq:vgve}), the task of finding the EM propagation in
periodic media is reduced to calculating the group velocity, i.~e.
$\vec{v}_g = \nabla_{\vec{K}}\omega$. This procedure may be called
the group velocity approach (GVA). As the group velocity is
relatively easy to be calculated from the band structures, this
method has been widely used.

From Eq.~(\ref{eq:ve1}), we see that the spatial average of the
current is taken over the whole area of the unit cell for two
dimensional cases. Experimentally, the energy flux is normally
determined by measuring currents flowing through certain surfaces.
In the following, we will examine whether and when the spatially
averaged currents can represent the actual flows. We will examine
a few averaging schemes.

\section{The systems and the various averaging schemes}

In the following, we will compare the energy flux obtained from
the GVA with the results from the direct computation of the
current given by Eq.~(\ref{eq:J}).

\subsection{The system}

The systems considered here are two dimensional photonic crystals
made of arrays of parallel dielectric cylinders placed in a
uniform medium, which we assume to be air. Such systems are common
in both theoretical simulations or experimental measurements of
two dimensional PCs\cite{Book}. For brevity, we only consider the
E-polarized waves (TM mode), that is, the electric field is kept
parallel to the cylinders. The following parameters are used in
the simulation. (1) The dielectric constant of the cylinders is
14, and the cylinders are arranged to form a square lattice. (2)
The lattice constant is $a$ and the radius of the cylinders is
0.3$a$; in the computation, all lengths are scaled by the lattice
constant. (3) The unit for the angular frequency is $2\pi c/a$.
After scaling, the systems become dimensionless; thus the features
discussed here would be applicable to a wider range of situations.

For the E-polarization , i.~e. the electrical field points along
the $z$ axis, the axis of the cylinders, Maxwell's equation is
simplified as \BE \left[\nabla^2 +
\epsilon(\vec{r})\omega^2/c^2\right]E(\vec{r})=0, \label{eq:me}
\EE where the electrical field is represented as \BE
\vec{E}(\vec{r}) = e^{-i\omega t}E(\vec{r})\hat{z}.\EE By Bloch's
theorem, we have\BE E(\vec{r}) =
e^{i\vec{K}\cdot\vec{r}}\sum_{\vec{G}}
E_{\vec{G}}e^{i\vec{G}\cdot\vec{r}}, \label{eq:bw}\EE where the
wave vector $\vec{K}$ is in the first Brillouin zone, and
$\vec{G}$ is a reciprocal vector. Substituting Eq.~(\ref{eq:bw})
into (\ref{eq:me}), we obtain the usual matrix equation \BE
\left[\omega^2/c^2 - (\vec{G} + \vec{K})^2\right ] E_{\vec{G}} = -
\sum_{\vec{G}'} \omega^2 B(\vec{G}-\vec{G}') E_{\vec{G}'},
\label{eq:band}\EE where $$B(\vec{G}-\vec{G}') = 2\pi(\epsilon
-1)a J_1(|\vec{G}-\vec{G}'|)/(|\vec{G}-\vec{G}'|d^2),$$ with $J_1$
being the Bessel function of the first order. After $E$ is
obtained, the magnetic field can be deduced by employing
Eq.~(\ref{eq:EM}). The energy current can be therefore derived
from Eq.~(\ref{eq:J}).

\subsection{Various averaging schemes}\label{schemes}

\begin{figure}[hbt]
\begin{center}
\epsfxsize=5in \epsffile{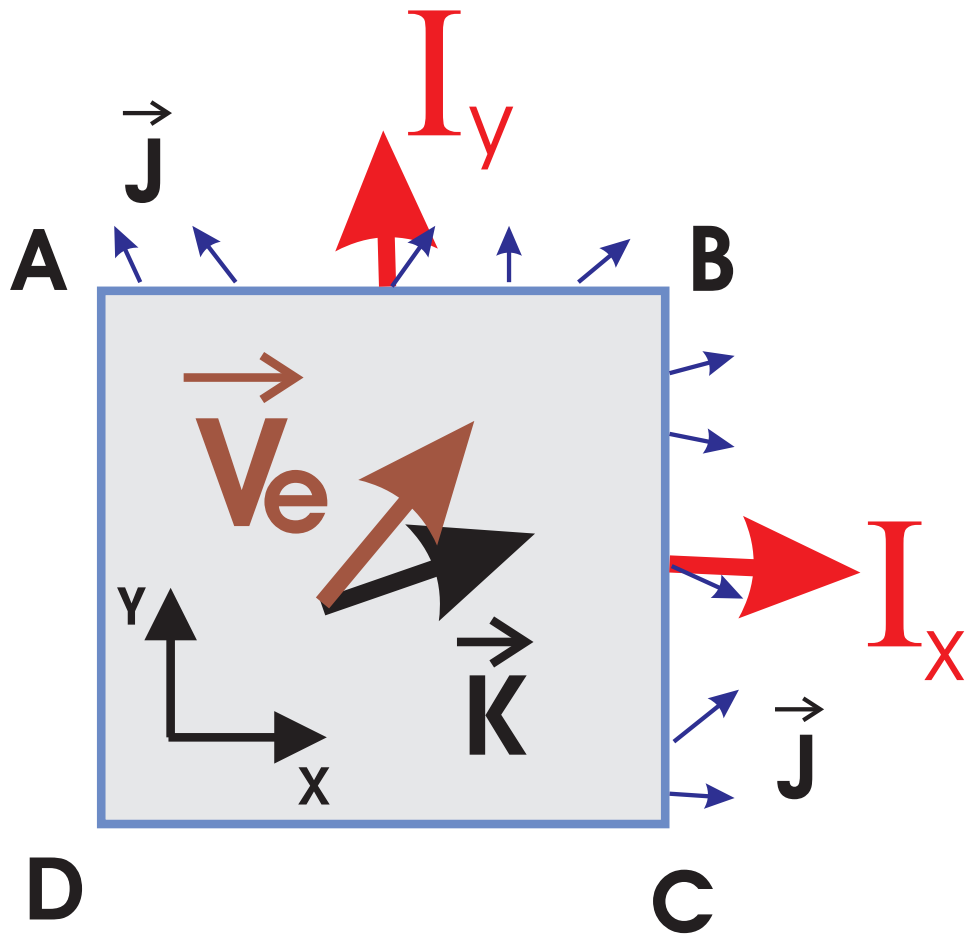} \caption{ \label{fig1}
Conceptual layout of the EM energy flow in a unit cell.
}\end{center}
\end{figure}

To compare the energy current determined from the GVA with that
obtained from the direct computation method, we take the following
procedure. Due to the periodicity and symmetry, it is sufficient
to just consider the energy current in a unit cell, which takes
the square shape. For a given frequency, a Block wave vector
$\vec{K}$ can be determined from the secular equation for the band
structure in Eq.~(\ref{eq:band}). The group velocity, therefore
the energy velocity, will then be calculated for $\vec{K}$ from
the band structure as \BE\vec{v}_g = \nabla_{\vec{K}}\omega.\EE
This consideration is illustrated by Fig.~\ref{fig1}. Here, the
coordinates are shown in the figure.

According to the aforementioned equivalency theorem, the direction
of the group velocity will be the direction of the spatially
averaged EM energy current, i.~e. \BE \langle\vec{J}\rangle =
\frac{1}{S}\int_S d^2\vec{r}\vec{J}. \label{eq:whole}\EE We note
that the integration is performed over the whole area of the unit
cell, which includes the areas occupied by the scatterers. In
actual experiments, it is often difficult to probe the currents
within the areas taken by the scatterers, therefore we may replace
the whole integration by a partial integration \BE
\langle\vec{J'}\rangle = \frac{1}{S'}\int_{S'} d^2\vec{r}\vec{J}.
\label{eq:partial}\EE Here the integration is performed over the
area $S'$ which excludes the areas occupied by the scatterers.

In practice, the EM energy current through this unit cell can also
be calculated from the flux across the two sides of the cell
denoted by AB and BC, i.~e. \BE I_y \sim \int_{AB} dx
\vec{J}\cdot\hat{y}, \ \mbox{and}\ I_x \sim \int_{BC} dy
\vec{J}\cdot\hat{x}. \label{eq:actual01}\EE Here $\vec{J}$ is
given by Eq.~(\ref{eq:J}).

We will compare the results in Eqs~(\ref{eq:whole}),
(\ref{eq:partial}) and (\ref{eq:actual01}) with that obtained in
Eq.~(\ref{eq:vgve}) by the GVA: \BE I_y^{GVA} \sim
(\nabla_{\vec{K}}\omega)\cdot\hat{y}, \ \mbox{and} \ I_x^{GVA}
\sim (\nabla_{\vec{K}}\omega)\cdot\hat{x}.\EE In the present
paper, we label the averaged current in Eq.~(\ref{eq:actual01}) as
case 1, that in Eq.~(\ref{eq:whole}) as case 2, and that in
Eq.~(\ref{eq:partial}) as case 3.

To simplify our discussion, we will compare the ratio between the
averaged current in two directions.
\begin{itemize}
\item The GVA case. The angle of the group velocity is determined
as \BE \phi_{GVA} =
\tan^{-1}\left(\frac{\nabla_{\vec{K}}\omega\cdot\hat{y}}{\nabla_{\vec{K}}\omega\cdot\hat{x}}\right).\label{eq:phig}\EE
\item Case 1: $\frac{I_y}{I_x}$ from Eq.~(\ref{eq:actual01}). we
represent the ratio by the angle \BE \phi_1 =
\tan^{-1}\left(\frac{I_y}{I_x}\right),\label{eq:phi1}\EE \item
Case 2: $\frac{\langle \vec{J}\rangle_y}{\langle
\vec{J}\rangle_x}$ from Eq.~(\ref{eq:whole}). The corresponding
angle is \BE \phi_2 = \tan^{-1}\left(\frac{\langle
\vec{J}\rangle_y}{\langle \vec{J} \rangle_x}\right).
\label{eq:phi2}\EE \item Case 3: $\frac{\langle
\vec{J}'\rangle_y}{\langle \vec{J}'\rangle_x}$ from
Eq.~(\ref{eq:partial}). The corresponding angle is \BE \phi_3 =
\tan^{-1}\left(\frac{\langle \vec{J}'\rangle_y}{\langle
\vec{J}'\rangle_x}\right).\label{eq:phi3}\EE
\end{itemize}

There are other options in obtain the averaged current. We refer
to the setup shown in Fig.~\ref{fig1}. If the detection is along
line AB, the averaged current vector may be obtained as \BE
\langle \vec{J}_{AB} \rangle = \frac{1}{L_{AB}}\int_A^B dx
\vec{J}. \label{eq:AB}\EE If the detection is on line BC, the
averaged current vector will be \BE \langle \vec{J}_{CB} \rangle =
\frac{1}{L_{CB}}\int_C^B dy \vec{J}. \label{eq:BC}\EE Here
$L_{AB}$ and $L_{CB}$ denote the lengths of the two sides of the
unit cell. We may also consider the sum of these two averaged
current if the detection is made on both AB and BC.
Correspondingly, there are other three possibilities
\begin{itemize}
 \item Case 4: $\frac{\langle \vec{J}_{CB}
\rangle_y}{\langle \vec{J}_{CB} \rangle_x}$ from
Eq.~(\ref{eq:BC}). The associated angle is \BE \phi_4 =
\tan^{-1}\left(\frac{\langle \vec{J}_{CB}\rangle_y}{\langle
\vec{J}_{CB}\rangle_x}\right). \label{eq:phi5}\EE \item Case 5:
$\frac{\langle \vec{J}_{AB} \rangle_y}{\langle \vec{J}_{AB}
\rangle_x}$ from Eq.~(\ref{eq:AB}). The associated angle is \BE
\phi_5 = \tan^{-1}\left(\frac{\langle
\vec{J}_{AB}\rangle_y}{\langle \vec{J}_{AB}\rangle_x}\right).
\label{eq:phi4}\EE
 \item Case 6:
$\frac{(\langle \vec{J}_{AB} \rangle + \langle
\vec{J}_{CB}\rangle)_y}{(\langle \vec{J}_{AB} \rangle + \langle
\vec{J}_{CB}\rangle)_x}$ from Eqs.~(\ref{eq:AB}) and
(\ref{eq:BC}). The associated angle is \BE \phi_6 =
\tan^{-1}\left(\frac{(\langle \vec{J}_{AB} \rangle + \langle
\vec{J}_{CB}\rangle)_y}{(\langle \vec{J}_{AB} \rangle + \langle
\vec{J}_{CB}\rangle)_x}\right). \label{eq:phi6} \EE
\end{itemize}

Since the energy or the current fields in the areas occupied by
the scatterers, may not be easy to detect, thus there is another
possibility in case 4, 5 and 6. That is, the contribution from
these areas are excluded. Later we will compare the results
obtained from various averageing schemes.

\section{The results and discussion}

\begin{figure}[hbt]
\begin{center}
\epsfxsize=5in \epsffile{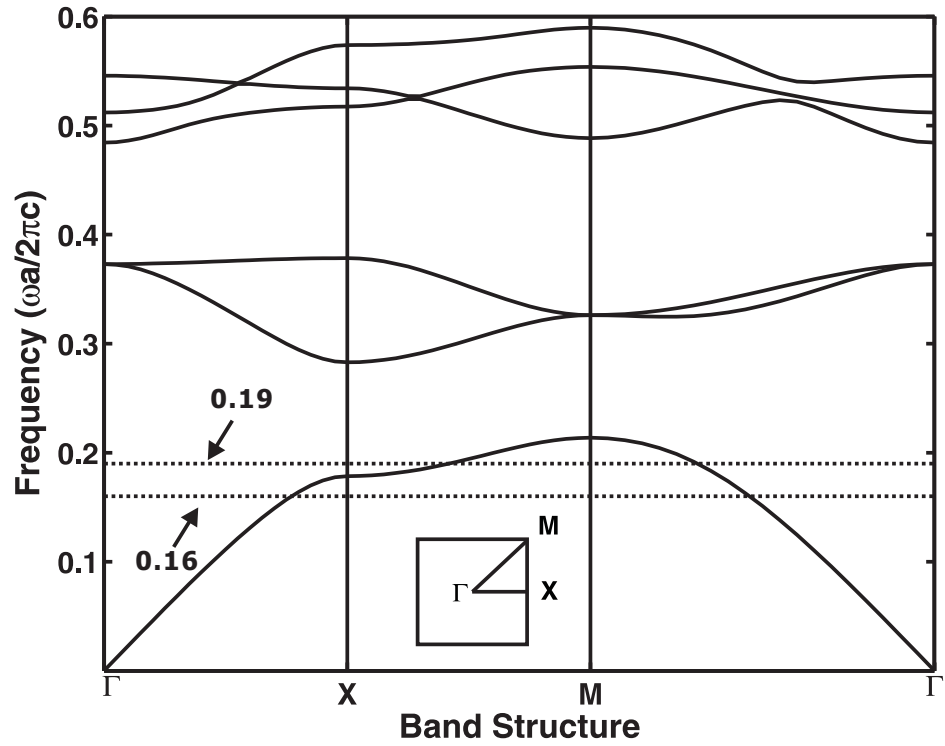} \caption{ \label{fig2} The band
structure of a square lattice of dielectric cylinders. The lattice
constant is $a$ and the radius of the cylinders is $0.3a$. $\Gamma
M$ and $\Gamma X$ denote the [11] and [10] directions
respectively.}\end{center}
\end{figure}

The frequency band structure is plotted in Fig.~\ref{fig2}. A
complete band gap is shown between frequencies of 0.22 and 0.28.
Just below the complete gap, there is a regime of partial band gap
in which waves are not allowed to travel along the $\Gamma X$ or
[10] direction. We will consider waves two frequencies: one is
 at 0.16 which is in the first passing band, and the other is at 0.19
 which is within the partial bandgap.

\subsection{Two dimensional imaging of energy and energy current
fields}

\begin{figure}[hbt]
\begin{center}
\epsfxsize=5in \epsffile{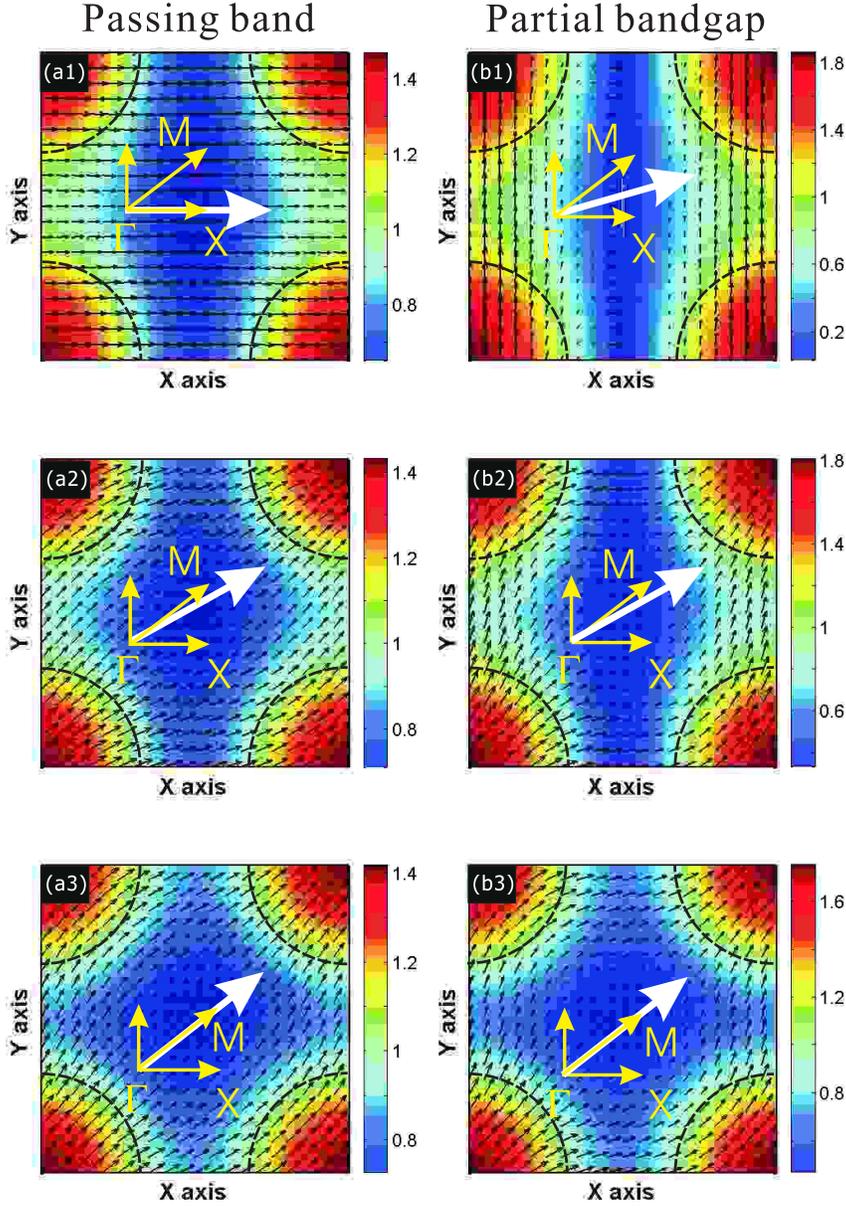} \caption{ \label{fig3} The
imaging of the intensity fields $|\vec{E}|$ and the current of
eigen-modes. Two frequencies are taken: 0.16 and 0.19 for the left
and right panel respectively. (a1) $\vec{K} = (0.77\pi/a,0)$, i.e.
the Bloch vector points to the angle of 0$^o$; (a2) $\vec{K} =
(0.66\pi/a,0.38\pi/a)$, i.e. the Bloch vector points to the angle
of 30$^o$; (a3) $\vec{K} = (0.54\pi/a,0.54\pi/a)$, i.e. the Bloch
vector points to the angle of 45$^o$. (b1) $\vec{K} =
(0.99\pi/a,0.42\pi/a)$, i.e. the Bloch vector points to the angle
of 23$^o$; (b2) $\vec{K} = (0.87\pi/a,0.51\pi/a)$, i.e. the Bloch
vector points to the angle of 30$^o$; (b3) $\vec{K} =
(0.69\pi/a,0.69\pi/a)$, i.e. the Bloch vector points to the angle
of 45$^o$.}\end{center}
\end{figure}

First we study the spatial behavior of the energy density fields
and the local flows of the eigen-modes which are characterized by
the Bloch wave vectors. The current is computed by
Eq.~(\ref{eq:J}), while the density field is calculated by
Eq.~(\ref{eq:U}). The results are shown in Figure~\ref{fig3}.

The left and right panel of Fig.~\ref{fig3} describes the result
for frequency 0.16 and 0.19 respectively. The former lies within
the first passing band, whereas 0.19 is within the first partial
bandgap regime. Within this partial band, the waves are forbidden
from propagation along the $\Gamma X$ direction, i.e. [10]. For
each frequency, we have considered three eigen modes represented
by three Bloch vectors which are given in the figure caption. The
two principle directions of the unit cell, $\Gamma X$ and $\Gamma
M$, are also shown in the figure.

Here we observe the following. First we discuss the case of the
passing band in (a1), (a2) and (a3). Overall speaking, the flows
of the energy indicated by the black arrows tend to flow along the
direction indicated by the Bloch vectors. This feature is more
obvious for the local current within the dielectric cylinders. For
the frequency within the partial gap, however, we observe that
most of the current flows may not be along the direction of the
Bloch vector. Fig.~\ref{fig3}(b1) shows that for small angles with
reference to the [10] direction, all the local currents tends to
flow along the direction which is close to the direction of [01].
That is, the energy flows are nearly vertical, but they cannot be
exactly vertical, as the direction of [01] is a forbidden
direction.

When the direction of the Bloch vector is tilted more and more
away from the direction [10], an interesting feature prevails.
That is, the currents eventually tends to flow into the direction
of $\Gamma M$, i.~e. the [11] direction. This feature is clearly
demonstrated by the examples in Fig.~3(b2) and (b3) for which the
Bloch vector points to the angles of 30$^o$ and 45$^o$
respectively.

Comparing the results for the passing band in the left panel and
the results for the partial band gap in right panel of
Fig.~\ref{fig3}, we may conclude that the electromagnetic flows in
periodic structures or photonic crystals will highly depend on the
band structures. There are significant differences in the current
behavior between the situation in which the frequency is located
in a passing band and the case in which the frequency is within a
partial band gap. The result shown by Fig.~\ref{fig3}(b2) suggests
that an effect of the partial band gap is to bend the current
towards a direction which is to avoid the forbidden direction as
much as possible; in the present case, it is the direction of
$\Gamma M$ or [11]. Such feature may render possible new
applications of partial band gaps in manipulating EM waves in
opto-electronic devices. The results in Fig.~\ref{fig3} also shows
that both the energy fields and the current fields are not uniform
inside the unit cell. This feature indicates that the energy
velocity is not also not uniform. The energy is more concentrated
within the regimes occupied by the scatterers.

\subsection{Comparison of different methods obtaining the
averaged currents}

The results in Fig.~\ref{fig3} indicate that the local energy
current is not uniform within a unit cell of a periodic structure.
How to describe the overall energy flow in such a non-uniform
situation of periodic structures thus poses an important issue.

As described in Section~\ref{GVA}, we mentioned that the common
theoretical approach is based on the equivalence theorem between
the group velocity and the averaged energy velocity\cite{Yariv}.
The average is taken over the whole volume in three dimensions or
the whole area in two dimensions, which include the volumes or the
areas occupied by the scatterers. In actual experiments or
observations, however, it may be difficult to probe the whole
volume or the whole area to deduce the information of the averaged
energy current. In particular, the currents or density within the
volumes or the areas occupied by the scatterers are hard to
detect. As matter of fact, there is no report on detecting the
energy or energy current over the whole volume or the area to our
best knowledge. The usual detection is made either at one
particular spatial point or on a certain surface. In addition, it
is often the intensity field that is measured.

In this section, we will compare the results obtained from the
various averaging schemes outlined in Section~\ref{schemes}. The
results will be compared with those obtained by the GVA. For
brevity yet without losing generality, we consider the two
frequencies from Fig.~\ref{fig3}: 0.16 and 0.19; one is in the
passing band and the other one is within the partial bandgap
regime.

\begin{figure}[hbt]
\begin{center}
\epsfxsize=5in \epsffile{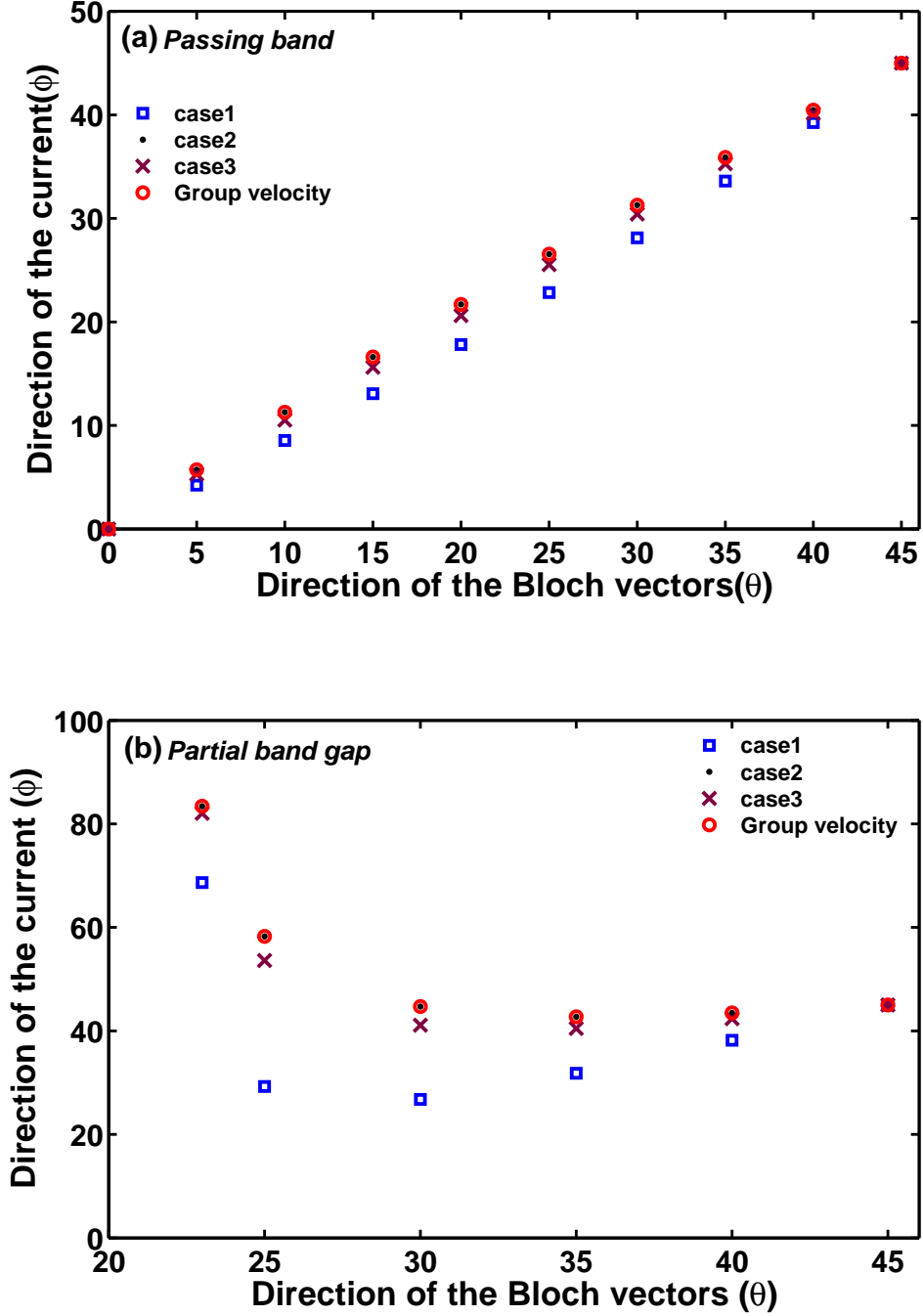} \caption{ \label{fig4}
Comparison of four different ways obtaining the averaged EM
current in a unit cell. The three cases are from
Eqs.~(\ref{eq:phi1}), (\ref{eq:phi2}) and (\ref{eq:phi3})
respectively. The results from the GVA are obtained from
Eq.~(\ref{eq:phig}).}\end{center}
\end{figure}

First, we compare the first three cases of averaging and the GVA.
The three scenarios are given by Eqs.~(\ref{eq:phi1}),
(\ref{eq:phi2}), and (\ref{eq:phi3}) respectively. The results are
shown in Fig.~\ref{fig4}. Here the directions of various averaged
currents and the group velocity are plotted against the direction
of the Bloch wave vectors. The directions are represented by the
angles of the corresponding vectors with reference to the $x$
axis.

For either the passing band or the partial bandgap cases, we see
that the results from the averaging scheme 2, that is the average
is taken over the whole area of the unit cell, fully agree with
the results from the GVA. This verifies the equivalence theorem.
It can be also seen that the results from the scheme 3 also agree
reasonably well with the GVA. This implies that as long as the
current inside a periodic medium can be measured, the group
velocity can be well deduced by the averaging scheme, no matter
whether the areas occupied by the scatterers are excluded or not.
The situation is noticeably different for the averaging scheme 1,
in which the information relies on the measurements along the two
boundaries of the unit cell. This scheme can reproduce the results
of the GVA considerably well for the passing band case in
Fig.~\ref{fig4}(a), but fails for most of the Bloch wave vectors
in the partial bandgap case in Fig.~\ref{fig4}(b). In the later
case, the agreement recovers as the angle approaches 45 degree,
i.~e. as the direction of the Bloch vector approach that of
$\Gamma M$.

\begin{figure}[hbt]
\begin{center}
\epsfxsize=5.5in \epsffile{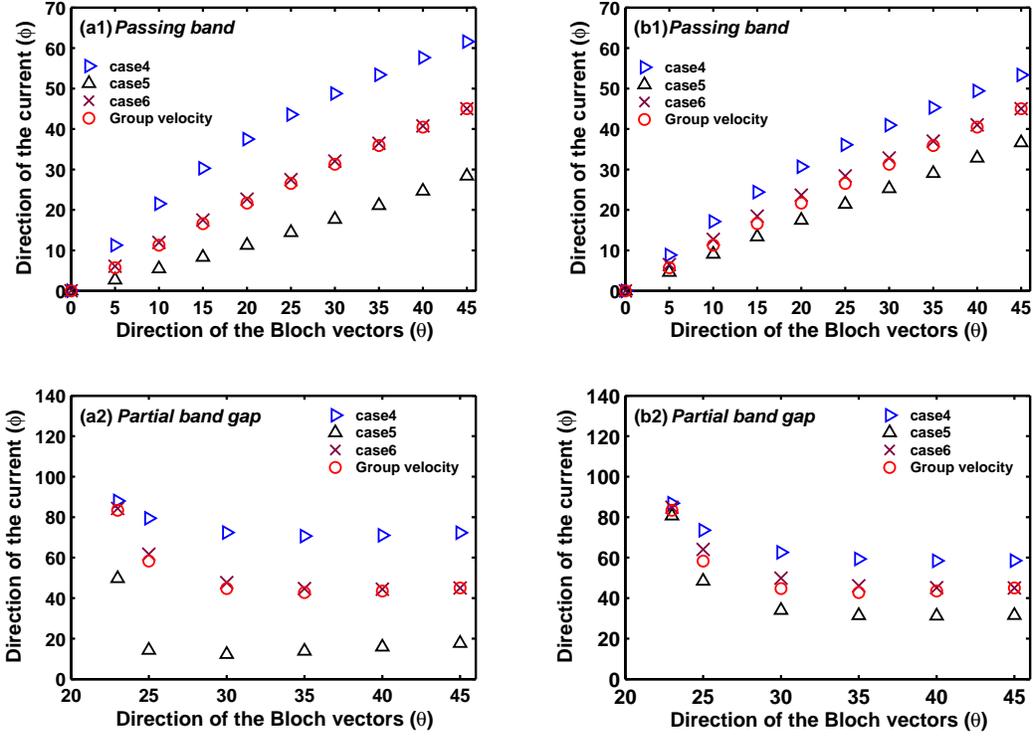} \caption{ \label{fig5}
Comparison of the results from the averaging schemes in case 4, 5
and 6. The results in the left panel are obtained as the areas
occupied by the cylinders are excluded in the average, whereas the
results in the right panel are obtained when the areas are
included. The three schemes are from Eqs.~(\ref{eq:phi4}),
(\ref{eq:phi5}) and (\ref{eq:phi6}) respectively. The results from
the GVA are obtained from Eq.~(\ref{eq:phig}). }\end{center}
\end{figure}

We have also compared the three other averaging schemes. The
results are presented in Fig.~\ref{fig5}. Here, we have considered
both the situation in which the areas occupied by the scatterers
(cylinders) are included and the case in which these areas are
excluded. Here we observe the following. (1) For both the passing
band and partial bandgap case, the average over any single side of
the photonic crystal will either overestimate (case 4) or
underestimate (case 5) the angle of the group velocity, no matter
whether the areas of the cylinders are included or not. In other
words, the group velocity will not represent the energy current
averaged only along one observation line, as represented by the
results in case 4 and 5. Relatively speaking, when the
contributions from the areas of cylinders are included, the
results move closer to that obtained from the GVA. (2) The results
from scheme 6 are in excellent agreement with the results from the
GVA, no matter whether the areas of the cylinders are included or
not. This feature implies that this averaging scheme is a good
candidate in inferring the group velocity of photonic crystals
when the energy density and energy current inside the crystals
cannot be readily probed.

Some common features can be discerned from Figs.~\ref{fig4} and
\ref{fig5}. In the passing band case, the direction of the group
velocity is close to that of the Bloch vector. The relation
between the angles of the two is almost linear, referring to
Fig.~\ref{fig4}(a) and Fig.~\ref{fig5}(a1) and (b1). For the first
partial bandgap situation, the angle of the group velocity
decreases as the angle of the Bloch vector $\vec{K}$ increases.
Actually as long as the angle of the Bloch eave vector exceeds 25
degree, the angle of the group velocity saturates to the value of
45 degree. This indicates that partial gaps tend to bend currents
into certain directions, allowing for possible novel applications
of photonic crystals in the partial bandgap regimes\cite{Chen}. We
have done further simulations, and found that these features are
also true for other frequencies in the first passing band and
first partial bandgap.

\subsection{A note on the energy density and energy flow in
photonic crystals}

From Fig.~\ref{fig3}, we see that the spatial distribution of the
energy density and the energy current is highly non-uniform in the
unit cell. This will also indicate that the local energy velocity,
defined as $\vec{v}_e \equiv
\frac{\vec{J}_{\vec{K}}}{U_{\vec{K}}}$ with $\vec{J}_{\vec{K}}$
and $U_{\vec{K}}$ being given in Eqs.~(\ref{eq:J}) and
(\ref{eq:U}), is also be non-uniform. This will have some
implications on the observation of the energy density and energy
flows. Since the current $\vec{J}$ equals $U\vec{v}_e$, then with
a given current magnitude, a larger local energy density
(intensity) implies a smaller velocity. For instance, consider two
current vectors, $\vec{J}_1$ and $\vec{J}_2$, with the same
magnitude, but perpendicular to each other. Clearly, the summation
of the two vectors, $\vec{J}_T = \vec{J}_1 + \vec{J}_2$, will give
a total vector which lies between the two vectors. If the
magnitudes of the two corresponding current velocities are not
equal, then the larger is the velocity magnitude, the smaller will
be the energy density. As a result, the apparent energy density
field will not be aligned along the direction of the total current
$\vec{J}_T$. This implies that the current flow may not be readily
obtainable by just measuring the energy intensity field. We may
also look at this problem from another perspective. The current
flow deduced from the group velocity approach, or the spatial
averaged current method in case 2 and 3 say, may not be able to
describe the apparent energy intensity field which is actually the
quantity to be measured.

\section{Summary}

We have considered the electromagnetic propagation in
two-dimensional periodic arrays of dielectric cylinders embedded a
uniform medium. The frequency band structure is computed using the
standard plane-wave expansion method, and the corresponding
eigne-modes are obtained subsequently. The spatially dependent
optical flows of the eigen-modes are calculated by a direct
computation approach. A few averaging schemes for the energy flows
are discussed. The results are compared to those obtained by the
common group velocity approach (GVA) which is based upon the group
velocity calculation. We have considered both the case in which
the frequency lies within passing band and the situation in which
the frequency is in the range of a partial bandgap. It is shown
that some average schemes may reproduce well the results of the
GVA. With these schemes, the group velocity can be deduced in
measurements. The research provides a useful information about how
to obtain the group velocity and what information the traditional
GVA can provide.

\section*{Acknowledgments}

The work received support from National Science Council, National
Central University.

%\newpage

\begin{appendix}

\section{Proof of $\vec{v}_e = \vec{v}_g$}

For the convenience of the reader, we present a proof of
$\vec{v}_e = \vec{v}_g$.

In photonic crystals, the waves assume the Bloch form, \BE
\vec{E}(\vec{r}) = \vec{E}_{\vec{K}}(\vec{r})
e^{i\vec{K}\cdot\vec{r}},\label{eq:BE} \EE and \BE
\vec{H}(\vec{r}) = \vec{H}_{\vec{K}}(\vec{r})
e^{i\vec{K}\cdot\vec{r}}, \label{eq:BH}\EE where $E_{\vec{K}}$ and
$\vec{H}_{\vec{K}}$ are periodic functions, and $\vec{K}$ is the
Bloch wave vector.

By substituting Eqs.~(\ref{eq:BE}) and (\ref{eq:BH}), we have \BE
\vec{v}_e = \frac{2\mbox{Re}[\langle\vec{E}_{\vec{K}}\times
\vec{H}^\star_{\vec{K}}\rangle]}{\langle\epsilon
|\vec{E}_{\vec{K}}|^2 + \mu|\vec{H}_{\vec{K}}|^2\rangle}, \EE
where $\langle\cdot\rangle$ denote the integration over the unit
cell.

Taking eqs.~(\ref{eq:BE}) and (\ref{eq:BH}) into the Maxwell
equations (\ref{eq:EM}), we have \BE \nabla \times
\vec{H}_{\vec{K}} + i\vec{K}\times \vec{H}_{\vec{K}} = -i\omega
\epsilon \vec{E}_{\vec{K}},\EE and \BE \nabla \times
\vec{E}_{\vec{K}} + i\vec{K}\times \vec{E}_{\vec{K}} = i\omega \mu
\vec{H}_{\vec{K}}.\EE

Suppose now that $\vec{K}$ is changed by an infinitesimal amount,
$\delta \vec{K}$. This will induce changes $\delta \omega$,
$\delta \vec{E}_{\vec{K}}$, and $\delta\vec{H}_{\vec{K}}$, and
they are related as \BE \nabla\times\delta\vec{H}_{\vec{K}} +
i\delta \vec{K}\times \vec{H}_{\vec{K}} + i \vec{K}\times
\delta\vec{H}_{\vec{K}} = -i\delta\omega\epsilon \vec{E}_{\vec{K}}
- i\omega\epsilon \delta\vec{E}_{\vec{K}} \EE and \BE
\nabla\times\delta\vec{E}_{\vec{K}} + i\delta \vec{K}\times
\vec{E}_{\vec{K}} + i \vec{K}\times \delta\vec{E}_{\vec{K}} =
i\delta\omega\mu \vec{H}_{\vec{K}} + i\omega\mu
\delta\vec{H}_{\vec{K}}. \EE In the following, let us ignore the
subscript $\vec{K}$.

Then we have \BE \nabla \times \vec{H} + i\vec{K}\times \vec{H} =
-i\omega \epsilon \vec{E},\EE and \BE \nabla \times \vec{E} +
i\vec{K}\times \vec{E} = i\omega \mu \vec{H},\EE and
 \BE \nabla\times\delta\vec{H} + i\delta \vec{K}\times \vec{H}
+ i \vec{K}\times \delta\vec{H} = -i\delta\omega\epsilon \vec{E} -
i\omega\epsilon \delta\vec{E}, \label{eq:DE} \EE and \BE
\nabla\times\delta\vec{E} + i\delta \vec{K}\times \vec{E} + i
\vec{K}\times \delta\vec{E} = i\delta\omega\mu \vec{H} +
i\omega\mu \delta\vec{H}. \label{eq:DH}\EE

Also we have \BE \nabla \times \vec{H}^* - i\vec{K}\times
\vec{H}^* = i\omega \epsilon \vec{E}^*, \label{eq:CM1}\EE and \BE
\nabla \times \vec{E}^* - i\vec{K}\times \vec{E}^* = -i\omega \mu
\vec{H}^*.\label{eq:CM2}\EE

Multiplying Eq.~(\ref{eq:DE}) with $\vec{E}^*$, we get
\begin{widetext} \BE \vec{E}^*\cdot(\nabla\times\delta\vec{H}) +
i\vec{E}^*\cdot(\delta\vec{K}\times \vec{H}) +
i\vec{E}^*\cdot(\vec{K}\times\delta\vec{H}) =
-i\delta\omega\epsilon|\vec{E}|^2 -
i\omega\epsilon\vec{E}^*\cdot\delta\vec{E}, \label{eq:1}\EE
\end{widetext}
and its complex conjugate is
\begin{widetext} \BE \vec{E}\cdot(\nabla\times\delta\vec{H}^*) +
i\vec{E}\cdot(\delta\vec{K}\times \vec{H}^*) +
i\vec{E}\cdot(\vec{K}\times\delta\vec{H}^*) =
-i\delta\omega\epsilon|\vec{E}|^2 -
i\omega\epsilon\vec{E}\cdot\delta\vec{E}^*, \label{eq:1a}\EE
\end{widetext}

Similarly, multiplying Eq.~(\ref{eq:DH}) with $\vec{H}^*$, we have
\begin{widetext} \BE \vec{H}^*\cdot(\nabla\times\delta\vec{E}) +
i\vec{H}^*\cdot(\delta\vec{K}\times \vec{E}) +
i\vec{H}^*\cdot(\vec{K}\times\delta\vec{E}) =
i\delta\omega\mu|\vec{H}|^2 +
i\omega\mu\vec{H}^*\cdot\delta\vec{H}, \label{eq:2}\EE
\end{widetext}
and its complex conjugate
\begin{widetext} \BE \vec{H}\cdot(\nabla\times\delta\vec{E}^*) +
i\vec{H}\cdot(\delta\vec{K}\times \vec{E}^*) +
i\vec{H}\cdot(\vec{K}\times\delta\vec{E}^*) =
i\delta\omega\mu|\vec{H}|^2 +
i\omega\mu\vec{H}\cdot\delta\vec{H}^*, \label{eq:2a}\EE
\end{widetext}

Subtracting Eq.~(\ref{eq:1}) from (\ref{eq:2}) and using \BE
\vec{A}\cdot(\vec{B}\times\vec{C}) =
\vec{B}\cdot(\vec{C}\cdot\vec{A}) =
\vec{C}\cdot(\vec{A}\times\vec{B}),\EE we get \begin{widetext}\BA
i\delta\omega(\epsilon |\vec{E}|^2 + \mu|\vec{H}|^2) +
i\omega(\epsilon\vec{E}^*\cdot\delta\vec{E} + \mu
\vec{H}^*\cdot\delta\vec{H}) &=&
i\delta\vec{K}\cdot(\vec{E}\times\vec{H}^* +
\vec{E}^*\times\vec{H}) \nonumber\\
+ \vec{H}^*\cdot[i \vec{K}\times\delta\vec{E} +
\nabla\times\delta\vec{E} ] -
\vec{E}^*\cdot[i\vec{K}\times\delta\vec{H} + \nabla\times\delta
\vec{H}] & &. \label{eq:3}\EA \end{widetext}

The last a few terms on the RHS of Eq.~(\ref{eq:3}) can be derived
as \begin{widetext}\BA \vec{H}^*\cdot[\nabla\times\delta\vec{E} +
i\vec{K}\times\delta\vec{E}] -
\vec{E}^*\cdot[i\vec{K}\times\delta\vec{H} + \nabla\times\delta
\vec{H}] &=& -\delta\vec{E}\cdot (i\vec{K}\times\vec{H}^*)
+\vec{H}^*\cdot(\nabla\times\delta\vec{E}) \nonumber \\
+ \delta\vec{H}\cdot(i\vec{K}\times\vec{E}^*) -
\vec{E}^*\cdot(\nabla\times\delta\vec{H}). && \label{eq:4}
\EA\end{widetext}

Now we use \BE i\vec{K}\times \vec{H}^* = \nabla \times \vec{H}^*
- i\omega \epsilon \vec{E}^*, \label{eq:CM3}\EE and \BE
i\vec{K}\times \vec{E}^* = \nabla\times \vec{E}^* + i\omega \mu
\vec{H}^*.\label{eq:CM4} \EE Then \begin{widetext}\BA I &=&
-\delta\vec{E}\cdot (i\vec{K}\times\vec{H}^*)
+\vec{H}^*\cdot(\nabla\times\delta\vec{E}) +
\delta\vec{H}\cdot(i\vec{K}\times\vec{E}^*) -
\vec{E}^*\cdot(\nabla\times\delta\vec{H}) \nonumber\\
&=& \ i\omega\epsilon \vec{E}^*\cdot\delta\vec{E}   - \
\delta\vec{E}\cdot(\nabla\times\vec{H}^*) +
\vec{H}^*\cdot(\nabla\times\delta\vec{E}) \nonumber\\
& & +\ i\omega\mu \vec{H}\cdot\delta\vec{H}^* + \
\delta\vec{H}\cdot(\nabla\times\vec{E}^*)
-\vec{E}^*\cdot(\nabla\times\delta\vec{H})  \nonumber\EA
\end{widetext}
or
\begin{widetext}\BA
I &=&  i\omega\epsilon \vec{E}^*\cdot\delta\vec{E} + i\omega\mu
\vec{H}\cdot\delta\vec{H}^* +
\nabla\cdot(\vec{H}^*\times\delta\vec{E}) +
\nabla\cdot(\delta\vec{H}\times\vec{E}^*). \EA\end{widetext}

Therefore Eq.~(\ref{eq:3}) becomes \begin{widetext}\BA
i\delta\omega(\epsilon |\vec{E}|^2 + \mu|\vec{H}|^2) +
i\omega(\epsilon\vec{E}^*\cdot\delta\vec{E} + \mu
\vec{H}^*\cdot\delta\vec{H}) &=&
i\delta\vec{K}\cdot(\vec{E}\times\vec{H}^* +
\vec{E}^*\times\vec{H}) + \nonumber\\
&& i\omega\epsilon \vec{E}^*\cdot\delta\vec{E} + i\omega\mu
\vec{H}\cdot\delta\vec{H}^* + \nonumber\\ &&
\nabla\cdot(\vec{H}^*\times\delta\vec{E}) +
\nabla\cdot(\delta\vec{H}\times\vec{E}^*).
\nonumber\\
\label{eq:5}\EA
\end{widetext}
Its complex conjugate is
\begin{widetext}\BA
-i\delta\omega(\epsilon |\vec{E}|^2 + \mu|\vec{H}|^2)
-i\omega(\epsilon\vec{E}\cdot\delta\vec{E}^* + \mu
\vec{H}\cdot\delta\vec{H}^*) &=&
-i\delta\vec{K}\cdot(\vec{E}^*\times\vec{H} +
\vec{E}\times\vec{H}^*) + \nonumber\\
&& -i\omega\epsilon \vec{E}\cdot\delta\vec{E}^* - i\omega\mu
\vec{H}^*\cdot\delta\vec{H} + \nonumber
\\ && \nabla\cdot(\vec{H}\times\delta\vec{E}^*) +
\nabla\cdot(\delta\vec{H}^*\times\vec{E}).
\nonumber\\
\label{eq:5c}\EA
\end{widetext}

Due to the periodicity, for any periodic function $\vec{A}$ we
have \BE \int dv \nabla\cdot \vec{A} = \int d\vec{S} \cdot \vec{A}
= 0.\EE

Now we subtract Eq.~(\ref{eq:5c}) from Eq.~(\ref{eq:5}), then we
perform the volume integration over a unit cell to get \BE
2i\delta\omega\langle(\epsilon|\vec{E}|^2 + \mu|\vec{H}|^2)\rangle
= 4i\delta\vec{K}
\cdot\mbox{Re}[\langle\vec{E}\times\vec{H}^*\rangle].\EE That is
\BE \delta\omega = \vec{v}_e\cdot\delta\vec{K}.\label{eq:ve}\EE

Comparison between Eqs.~(\ref{eq:vg}) and (\ref{eq:ve}) leads to
\BE \vec{v}_g = \vec{v}_e.\EE

From the above derivation, we understand the conditions for
$\vec{v}_g = \vec{v}_e$ to be held are that (1) the variation
$\delta \vec{K}$ can be arbitrary; (2) when the variation $\delta
\vec{K}$ is small, the changes in $\omega$, $E_K$ and $H_K$ are
also small.

\end{appendix}

\newpage

\newpage

\newpage

\newpage

\newpage

\end{document}